\documentclass[10pt]{article} 
\usepackage[superscript]{cite}
\usepackage{amsmath}
\usepackage{graphics}
\usepackage{indentfirst}

\newcommand{\onlinecite}[1]{\citen{#1}}
\newcommand{\ocite}[1]{\onlinecite{#1}}
\newcommand{\rcite}[1]{Ref.~\onlinecite{#1}}
\newcommand{\rrcite}[1]{Refs.~\onlinecite{#1}}

\newcommand{\ad}{^\dagger}

\newcommand{\AND}{{\small AND}}

\newcommand{\becs}{\begin{cases}}
\newcommand{\Blp}{\Big(}
\newcommand{\Brp}{\Big)}

\newcommand{\dya}[1]{|#1\rangle\langle#1|}
\newcommand{\inp}[1]{\langle#1|#1\rangle}
\newcommand{\inpd}[2]{\langle#1|#2\rangle}
\newcommand{\ket}[1]{|#1\rangle}

\newcommand{\mte}[2]{\langle#1|#2|#1\rangle}
\newcommand{\od}{\odot}
\newcommand{\OR}{{\small OR}}
\newcommand{\ot}{\otimes}

\newcommand{\ra}{\rightarrow}

\newcommand{\st}{\sqrt{2}}

\newcommand{\tm}{\times}

\newcommand{\vb}{\,|\,}

\newcommand{\HM}{{\mathcal H}}

\newcommand{\PM}{{\mathcal P}}
\newcommand{\QM}{{\mathcal Q}}

\newcommand{\al}{\alpha}
\newcommand{\bt}{\beta}

\newcommand{\dl}{\delta}

\renewcommand{\th}{\theta} 

\begin{document}

\title{EPR, Bell, and quantum locality}

\author{Robert B. Griffiths%
\thanks{Electronic mail: rgrif@andrew.cmu.edu}\\
Department of Physics,
Carnegie-Mellon University,\\
Pittsburgh, PA 15213, USA}

\date{Version of 22 July 2011}

\maketitle  

\begin{abstract}
  Maudlin has claimed 
that no local theory can reproduce the predictions of
  standard quantum mechanics that violate Bell's inequality for Bohm's version
  (two spin-half particles in a singlet state) of the Einstein-Podolsky-Rosen
  problem. It is argued that, on the contrary, standard quantum mechanics
  itself is a counterexample to Maudlin's claim, because it is local in the
  appropriate sense (measurements at one place do not influence what occurs
  elsewhere there) when formulated using consistent principles in place of the
  inconsistent appeals to ``measurement'' found in current textbooks. This
  argument sheds light on the claim of Blaylock that counterfactual
  definiteness is an essential ingredient in derivations of Bell's inequality.
\end{abstract}

\tableofcontents

\section{Introduction}
\label{sct1}

Blaylock has argued that derivations of Bell's inequality involve an implicit
assumption of counterfactual definiteness as well as locality, and
consequently the fact that quantum theory violates Bell's inequality is
consistent with the quantum world being local (as well as realistic and
deterministic).\cite{Blyl10} In response Maudlin\cite{Mdln10} asserted that
Bell's result does not require counterfactual definiteness, and the fact that
quantum theory violates it means that the quantum world is nonlocal: no local
theory can reproduce The Predictions of quantum mechanics for the results of
experiments done very far apart.

In this article we shall show that this last assertion is mistaken, by
exhibiting a local theory for the situation of interest: measurements of
various components of spin angular momentum on two spin-half particles
prepared in a singlet state.  (Blaylock considers polarization measurements on
two photons, but these measurements amount to the same situation for the
issues under discussion.) Because Maudlin's argument is clearly written and
relatively short, it is possible to make a useful comparison between it and
the sort of reasoning needed to yield consistent results in the quantum
domain. The connection with counterfactual definiteness will also be
discussed.

The reader is probably aware that Bell's inequality --- see \rcite{Blyl10} for
extensive references --- puts constraints on certain correlation functions
relating properties of separated systems, or measurements of these properties.
Quantum mechanics predicts violations of these constraints, and measurements
confirm the predictions of quantum theory. Consequently, any derivation of
Bell's inequality must violate one or more of the fundamental principles of
quantum mechanics; or to put it another way, when viewed from a quantum
perspective
such a derivation contains one or more mistakes. On this point Blaylock,
Maudlin, and the present author are in full agreement. The issue is: where and
what are the mistakes? Finding mistakes is sometimes difficult. Many of us
have experienced the frustration of dealing with a student asking us to find
the error in a lengthy argument which he thinks is valid, but which leads to a
result that we know is wrong. Various derivations of Bell's inequality are
complicated and involve probabilistic reasoning that is not
straightforward. And sometimes additional assumptions, such as the existence
of agents with free choice,\cite{Stpp97,dEsp06} move the discussion out of the
realm of simple physics. Understanding these arguments, much less finding
mistakes in them, can be an arduous task. We are fortunate that the argument
for nonlocality in \rcite{Mdln10} is clear, compact, well written, and free
choice plays no essential role.

The present paper is not intended to discuss all the assumptions that underlie
various derivations of Bell's inequality, or their possible validity in
domains unrelated to quantum theory. Instead, the primary issue addressed is:
at what point do Bell's arguments, in particular as presented in
\rcite{Mdln10}, conflict with quantum theory?

Maudlin is not alone in maintaining that violations of Bell's inequality imply
that quantum mechanics is nonlocal. For an extended, though by no means
complete, list of publications that have maintained this position, together
with a significant number that have expressed disagreement, see the
bibliography in \rcite{Grff11}, which contains a more detailed analysis of the
Bell inequality from the perspective summarized in the present paper, along
with a proof of an appropriate form of Einstein locality.

Issues of the sort under discussion cannot be resolved within the framework of
standard quantum mechanics, when that is understood to be what is found in
contemporary textbooks.  Although textbooks provide students with many
effective and efficient techniques for doing calculations, they do not contain
an adequate presentation of the fundamental principles of the theory needed to
understand why these techniques lead to reasonable answers, and what are their
limits of applicability. Students become skilled at calculating the right
answer, but are left confused or uncertain as to why, or whether, the answer
really is right. Inquiries from the perplexed are met with
Mermin's\cite{Mrmn89} well-known dictum: Shut up and calculate! In particular,
attempts to make \emph{measurements} part of the foundation of quantum
mechanics are unsatisfactory because of the infamous \emph{measurement
  problem}: the inability of such an approach to incorporate a physical
measuring apparatus in fully quantum mechanical terms. The difficulties were
pointed out by Wigner,\cite{Wgnr63} and remain unsolved.\cite{Mttl98,BsSh96}

Thus in order to address the question of whether quantum theory is local, a
consistent formulation of the principles that lie behind, and can be used to
justify, the textbook calculations is needed; a formulation that, as Bell
emphasized,\cite{Bll90} is \emph{not} based upon measurements, but instead
allows us to understand, in quantum-mechanical terms, what goes on in real
measuring processes. Such a formulation, in which probabilities (but not
measurements) are fundamental, exists, and deserves to be better known. We
shall refer to it as the \emph{histories} approach; alternative names are
\emph{consistent histories}\cite{Grff84} and \emph{decoherent histories}.
\cite{GMHr90} For further references and details see \rcite{Grff02c}; for
short introductions see \rrcite{Grff02b} and \ocite{Hhnb10}. It yields what
Maudlin calls ``The Predictions'' for Bohm's version\cite{Bhm51s} of the
Einstein-Podolsky-Rosen (EPR) problem.\cite{EnPR35} It is explicitly local, as
discussed below, and was applied by the author to the EPR problem
in Ref.~\onlinecite{Grff87}.

Because many readers will not be familiar with the histories approach (it is
not mentioned in Refs.~\onlinecite{Blyl10} and \onlinecite{Mdln10}), some of
its fundamental features are presented in a compact form in Sec.~\ref{sct2},
which deals with quantum statics, and in Sec.~\ref{sct3}, devoted to quantum
dynamics. The histories approach is not based upon the concept of measurement,
and readers who try to understand it in terms of measurements are likely to
become confused. It is better to start with a mental picture of a classical
system inside a closed box in which the dynamics is intrinsically random or
stochastic. Things actually happen inside the box, but what happens in the
future (or in the past) cannot be inferred with certainty from the state of
affairs at a given point in time. As with any classical picture of the world,
this one can only be a first step on the way to a fully consistent quantum
description, but it is less likely to mislead than trying to think about
things in terms of measurements associated with mysterious wave function
collapses, as in current textbooks.

Section~\ref{sct4} discusses how genuine measurements, understood as real
physical processes that go on in the real (quantum) world, can be understood
from the perspective of fundamental quantum theory. The tools will then be in
hand for a consistent analysis in Sec.~\ref{sct5} of the Bohm version of the
EPR situation using two spin-half particles in a singlet state, both with and
without measurements, demonstrating the locality of quantum theory.

Maudlin's argument for nonlocality\cite{Mdln10} is examined
in Sec.~\ref{sbct6.1} in order to locate how and where it diverges from the
analysis in Sec.~\ref{sct5}. Counterfactuals in the quantum domain are
discussed in Sec.~\ref{sbct6.2}, with the conclusion that their misuse can
plausibly be part of one route leading to Bell's inequality, as Blaylock
indicates in \rcite{Blyl10}, though perhaps it is not the only route.
The paper ends with a brief summary and some comments in Sec.~\ref{sct7}.

\section{Quantum statics}
\label{sct2}

The most fundamental difference between classical and quantum mechanics is
that the former makes use of a phase space whose individual points represent
possible states of a physical system, with subsets of points representing
physical properties.  Whereas, as explained by von Neumann \cite{vNmn55b},
quantum mechanics uses a complex Hilbert space, and physical properties
correspond to subspaces, with a one-dimensional subspace (ray) the quantum
analog of a single point in phase space.  A finite-dimensional Hilbert space
is adequate for purposes of the following discussion.  In classical mechanics
the logical negation of the property $P$ that corresponds to a set of points
$\PM$ in the phase space, which is to say the property ``not $P$'', is
represented by the set-theoretic complement $\PM^c$ of the set $\PM$.  In the
quantum case negation\cite{vNmn55b} corresponds to the \emph{orthogonal
  complement} $\PM^\perp$ of the subspace $\PM$: the collection of all vectors
in the Hilbert space that are orthogonal to every vector in the
subspace. Equivalently, if $P$ is the projector onto $\PM$, its negation is
represented by the projector $I-P$, where $I$ is the identity operator on the
Hilbert space.

\begin{figure}[h]
$$\includegraphics{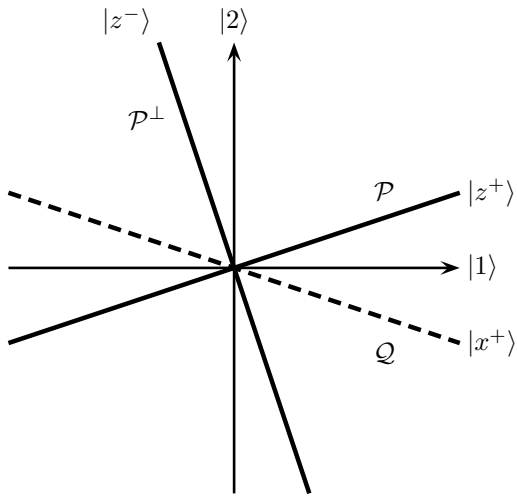}$$
\caption{Schematic diagram of a two-dimensional Hilbert space using the real
plane, with a basis consisting of the orthogonal kets $\ket{1}$ and $\ket{2}$,
and various rays (one-dimensional subspaces) as discussed in the text.}
\end{figure}

This leads to a profound difference between classical and quantum properties,
which is best illustrated by considering a two-dimensional Hilbert space
representing possible properties of a spin-half particle, shown schematically
in Fig.~1. (The figure shows a real, rather than a complex Hilbert space, but
it is adequate for present purposes.) The line $\PM$ through the origin
represents the physical property $S_z=+1/2$ in units of $\hbar$. The line
$\PM^\perp$ perpendicular to this line is its orthogonal complement and
represents the physical property $S_z=-1/2$. The key point is that there are
many other lines through the origin, such as $\QM$ corresponding
(schematically) to the property $S_x=1/2$, which are neither the same as the
property $S_z=+1/2$ or its negation, $S_z=-1/2$.  This is fundamentally
different from classical phase space where for a given property $P$, any point
in the phase space is either inside the set $\PM$, which means the system this
property $P$, or in its complement $\PM^c$, which means the system does not
have property $P$. In quantum mechanics we say that two properties $P$ and
$Q$, such as $S_z=+1/2$ and $S_x=1/2$, are \emph{incompatible} when the
projectors onto the corresponding subspaces do not commute, $PQ\neq
QP$. Whenever noncommuting operators appear in quantum mechanics we have a
situation that lacks a precise analog in classical physics.

A consequence of adopting the von Neumann approach to quantum properties,
which underlies the presentations in all contemporary textbooks, is that
quantum reasoning must follow different rules from classical reasoning. For a
classical phase space the conjunction of two properties $P$ and $Q$, $P\land
Q$ or ``$P$ \AND\ $Q$,'' is always defined: it is the property corresponding
to the intersection $\PM\cap \QM$ of the corresponding sets of points. But in
the quantum case let $P$ be the property $S_z=+1/2$ and $Q$ be the
incompatible property $S_x=-1/2$. Can we make any sense of the conjunction
``$P$ \AND\ $Q$''? Does it correspond to any ray in the Hilbert space? The
answer is that it does not, or at least there is no plausible way to make such
an association. Every ray in Hilbert space can be interpreted as the property
that $S_w=+1/2$ for some direction in space $w$, so there is nothing left over
to represent a conjunction of the sort in which we are interested. There is no
room for it in Hilbert space.  What shall we
do?

Von Neumann was not unaware of this problem, and he and Birkhoff made a
proposal\cite{BrvN36} that if $P$ and $Q$ refer to incompatible properties,
``$P$ \AND\ $Q$'' should be the property corresponding to the
\emph{intersection} $\PM\cap\QM$ of the corresponding subspaces of Hilbert
space. As applied to the situation at hand, the intersection of the $S_z=+1/2$
and $S_x=-1/2$ rays is the subspace consisting of nothing but the zero vector,
the origin in Fig.~1. The orthogonal complement of the zero vector is the
entire Hilbert space, so the zero vector represents a proposition which is
\emph{always false}; it is the counterpart of the empty set for a classical
phase space.  This is an interesting proposal; let us pursue it for a bit and
see where it leads, using a notation in which single letters
\begin{equation}
z^+=\dya{z^+},\quad x^-=\dya{x^-},
\label{eqn1}
\end{equation}
and so forth correspond to projectors onto the rays $S_z=+1/2$, $S_x=-1/2$,
and so forth. Then the usual logical rules, with
$\land$ and $\lor$ meaning \AND\ and \OR, lead to the expression
\begin{equation}
\Blp z^+\land x^-\Brp\lor \Blp z^+\land x^+\Brp
= z^+ \land\Blp x^-\lor x^+\Brp = z^+,
\label{eqn2}
\end{equation}
where we have used the fact that $x^-\lor x^+$ is always true, because it is
the negation of the always false property $x^+\land x^-$.

But there is a serious problem with Eq.~\eqref{eqn2}. The proposition or
property on the left side is always false, as it is the disjunction of two
always-false properties.  So if we believe the equation, the property $z^+$,
the final term on the right, is always false.  We could just as well have used
$z^-$ in place of $z^+$, and drawn the conclusion that $z^-$ is always
false. Hence $z^+\lor z^-$ is always false. But this contradicts the
fact---see previous paragraph---that $z^+\lor z^-$ is always true! We have
arrived at a logical contradiction. How could two of the great mathematicians
of the 20th century have made such a blunder? The answer is that they did not.
They were quite aware that this problem was going to arise, and that the
ordinary rules of propositional logic would have to be modified. So they
proposed discarding the distributive rule, which justifies the first equality
in Eq.~\eqref{eqn2}, to produce what has come to be known as \emph{quantum
  logic}.

Does quantum logic provide a consistent underpinning of the calculational
techniques found in quantum textbooks? Not likely. Although quantum logic has
been extensively studied as a subdiscipline of quantum foundations, so far as
I am aware it has made no contribution to resolving the conceptual
difficulties besetting quantum theory.  It seems not to be used in quantum
information theory, which is currently the ``cutting edge'' of developments in
quantum theory, that is, in extending its concepts. This situation may simply
reflect the fact that professional physicists are not smart enough to really
understand quantum mechanics, and the problems of quantum foundations will
have to remain unresolved until robots surpass our intelligence and can learn
to think using the new rules. (But if they do come to understand quantum
mechanics, will they be smart enough, or even sufficiently motivated, to
explain it to \emph{us}?)

But surely the textbooks must do something with incompatible properties, and
if Birkhoff's and von Neumann's reasoning has not been adopted, what takes its
place? One can discern two strategies.  The first is based on measurements.
There is no way to simultaneously \emph{measure} $S_z$ and $S_x$ for a
spin-half particle.  On this everyone agrees, but we are back to the unsolved
measurement problem.  The second is to invoke the uncertainty principle.
There are elegant mathematical formulations of various uncertainty principles
to be found inside and outside textbooks.  But what do they mean?  What it
means in practice for the student learning the subject is that certain things
are best not discussed, or if they are, the discussion should be accompanied
by an appropriate amount of vagueness and arm waving.  If one is discussing
$S_z$, then it is probably best to leave $S_x$ out of the discussion, though
one is somewhat uncertain as to why certain things are uncertain.

The histories approach uses a precise principle in place of arm waving.
Incompatible properties \emph{cannot be combined} in a meaningful quantum
description of the world, even if the individual properties by themselves make
perfectly good sense. The statement ``$S_z=+1/2$ \AND\ $S_x=-1/2$'' makes no
sense. It is very important to keep in mind the distinction between ``makes no
sense'' or ``meaningless,'' and ``always false.'' A statement that is
meaningful and false --- see our discussion of the Birkhoff and von Neumann
approach --- is one whose negation is meaningful and true.  By contrast, the
negation of a meaningless statement is equally meaningless. In ordinary logic
the statement $P\land\lor\, Q$, which is to say ``$P$ \AND\ \OR\ $Q$,'' is
meaningless, because it has not been formed according to the syntactical rules
appropriate to the language. One should think of ``$S_z=+1/2$ \AND\
$S_x=-1/2$'' as similar -- it is meaningless in the sense that quantum
mechanics cannot assign it a meaning.

Similarly, ``$S_z=+1/2$ \OR\ $S_x=-1/2$'' is not a meaningful statement about
quantum properties, so it makes no sense to ask, ``Is the spin-half particle
in the state $S_x=1/2$ or the state $S_z=1/2$ or possibly both?'' Students are
taught that there is no way of measuring both $S_x$ and $S_z$ simultaneously,
and this statement is correct. Unfortunately, they are not taught that the
reason such measurements are impossible is that there is nothing there to be
measured: there is no physical property in the quantum world that such a
measurement might reveal. It is not the limited ability of experimental
physicists that is at issue; in fact, one sign of a good experimental
physicist is his inability to measure what is not there.

In the histories approach the rule for not combining incompatible properties is
part of a more general single framework rule. A \emph{framework} is a
projective decomposition of the identity: a collection of projectors
$\{P^\mu\}$, where $\mu$ is a label not an exponent, which sum to the identity
operator:
\begin{equation}
I= \sum_\mu P^\mu,\quad P^\mu = (P^\mu)\ad = (P^\mu)^2,\quad
P^\mu P^\nu = \dl_{\mu\nu} P^\mu.
\label{eqn3}
\end{equation}

Each projector corresponds to a physical property. The significance of $P^\mu
P^\nu =0$ when $\mu\neq\nu$ is that these properties are mutually exclusive:
if one is true the other is necessarily false. The collection of such
properties form a \emph{quantum sample space}: not only are they mutually
exclusive, but together they exhaust all possibilities --- that is the
significance of $\sum_\mu P^\mu=I$. The \emph{events}, using the terminology
of probability theory, associated with this sample space are projectors formed
by taking sums of elements in the collection $\{P^\mu\}$, including $I$ and
the 0 projector; together they comprise a Boolean algebra under (operator)
products and taking the complement $I-P$ of $P$.

Two such frameworks are \emph{compatible} if the projectors in the first
commute with all the projectors in the second; otherwise they are
\emph{incompatible}. Two compatible frameworks have a \emph{common
  refinement}: a single framework whose Boolean algebra includes all the
projectors of both of the original frameworks. Two incompatible frameworks
have no common refinement. The \emph{single framework rule} asserts that all
reasoning about a quantum system must be carried out in a single framework in
the sense that results from two frameworks \emph{cannot be combined}. Given
two compatible frameworks the single framework rule is easily satisfied: carry
out the reasoning using a common refinement.  Consequently, what the single
framework rule prohibits is combining, or attempting to combine, results from
incompatible frameworks.

It is important to understand just what the single framework rule allows and
what it prohibits. Physicists are free to employ whatever framework they wish
when describing a quantum system: formulas can be written down and
probabilities calculated using the $S_x$ or $\{x^+,x^-\}$, or the $S_z$
framework for a spin-half particle.  Call this freedom the principle of
Liberty. There is no law of nature, or of quantum theory, that says that one
of these is the ``right'' framework; from the perspective of fundamental
quantum theory they are equally correct. Call this the principle of
Equality. But Liberty and Equality are accompanied by a third principle:
Incompatibility. Results in two incompatible frameworks cannot be combined in
a meaningful way. What these principles amount to in practice will become
apparent when we consider various examples.

Note that when a physicist chooses to use a particular framework rather than
some other to describe a quantum system, this is not at all a matter of
somehow ``influencing'' the system in question. Instead, such choices are
somewhat like the choices a photographer makes in photographing Mount Rainier
from, say, the north rather than from the south. Such a choice determines what
kind of information is present, and consequently the utility or the use which
can be made of the resulting photograph. But it has no influence on the
mountain itself. Choosing a framework is thus very different from carrying out
a measurement on a system, because a measurement can have a rather drastic
effect upon a microscopic quantum system.

\section{Quantum dynamics}
\label{sct3}

Von Neumann\cite{vNmn55b} taught us that quantum dynamics involves two
distinct processes: a unitary or deterministic time evolution, and then a
separate stochastic or probabilistic time evolution associated in some way
with measurements. Although this is what students learn in textbooks, both
students and teachers (and, one suspects, textbook writers --- see
\rcite{Lloe01}) find it unaesthetic. One can sympathize with the efforts of
Everett and his successors\cite{Evrt57,DWGr73} to reduce all of quantum
dynamics to a unitary form (for a closed system), even though the ``many
worlds'' (or ``many minds'') approach has its own share of obscurities and
difficulties.\cite{Brrt99,Vdmn09,Brrt09} Blaylock\cite{Blyl10} is in favor of
the many-worlds interpretation, and Maudlin\cite{Mdln10} is not.

The histories approach used in this paper is the exact opposite of Everett's
in the sense that quantum time evolution is \emph{always} stochastic, or more
precisely, should always be put in a stochastic or probabilistic
framework. Classical (Hamiltonian) dynamics is deterministic: there is a
unique map carrying each point in the phase space at a give time to another
specific point at any particular later (or earlier) time. But there is no
reason why the quantum world should necessarily possess the same sort of
deterministic time development. And there is good experimental support, as in
the random time of decay of unstable particles, for the idea that quantum
dynamics has an intrinsically probabilistic character. But what about
\emph{the wave function} that satisfies the Schr\"odinger's (time-dependent
and deterministic) equation?  \emph{Solutions to Schr\"odinger's equation are
  to be used for calculating probabilities.}  This is what Born\cite{Brn26}
taught us and what is done in textbooks. Formulating probabilistic ideas using
histories, as in \rcite{Grff02c}, Chap.~8, results in a consistent approach
that removes (or at least ``tames'') quantum paradoxes.

A consistent discussion of a stochastic process, whether classical or quantum,
begins by introducing a sample space of mutually exclusive possibilities,
which we shall refer to as \emph{histories}. These can be represented, just as
quantum properties at a single time, by subspaces on a \emph{history} Hilbert
space consisting of a tensor product of copies of the Hilbert space describing
a quantum system at a single time, and with the projectors on the history
space playing a role analogous to those corresponding to properties at a
single time. (For details, see \rcite{Grff02c}, Chap.~8.) Once again, there
are many possibilities for such sample spaces, and the single framework rule
says that the physicist is welcome to use any one he pleases (Liberty
and Equality), as long as he does not put together meaningless combinations of
incompatible sample spaces (Incompatibility). The simplest sorts of histories
consist of just a sequence of quantum properties at a succession of different
times, and the set of properties considered at one time may be incompatible
with the set considered at a different time. Thus there is nothing inherently
wrong with saying that at successive times $t_1<t_2<t_3$ a spin-half particle
has $S_z=+1/2$ and then $S_x= -1/2$, and then $S_z = -1/2$, or perhaps
$S_y=-1/2$ at the third time. These properties need not, and generally do not,
map into one another under the unitary time evolution of a closed system
generated by Schr\"odinger's equation.

For this paper it will suffice to consider histories for a sequence of times
$t_0<t_1<t_2<\cdots t_f$ which can be represented in the form
\begin{equation}
Y^\al = \Psi_0\od P_1^{\al_1}\od P_2^{\al_2}\od\cdots P_f^{\al_f},
\label{eqn4}
\end{equation}
where $\Psi_0=\dya{\Psi_0}$ projects onto the initial state $\ket{\Psi_0}$ at
$t_0$, so all histories begin with this initial state, and $\{P_j^{\al_j}\}$
denotes a (projective) decomposition of the identity $I$ at time $t_j$, see
Eq.~\eqref{eqn3}. The time is labeled by a subscript, and decompositions at
different times need have no relation with one another. The superscript
on the history $Y^\al$ denotes a string of labels:
\begin{equation}
\al = (\al_1,\al_2,\ldots \al_f).
\label{eqn5}
\end{equation}
The $\od$ symbols in Eq.~\eqref{eqn4} can be interpreted as variants of the
usual $\ot$ symbol denoting a tensor product, but for present purposes we can
think of them as simply separating possible events at one time from events at
the next time.

Probabilities can be assigned to histories if such a family represents events
in a closed quantum system, in the following way. Let $T(t',t)$ be the unitary
\emph{time development} operator; for example, if the system Hamiltonian $H$
is independent of time, then $T(t',t)=\exp[-i(t'-t) H/\hbar]$. For history
$\al$ define the corresponding \emph{chain ket}:
\begin{equation}
\ket{\al} = P_f^{\al_f} T(t_f,t_{f-1})\cdots P_1^{\al_1} T(t_1,t_0)\ket{\Psi_0}.
\label{eqn6}
\end{equation}
Then provided the \emph{consistency condition}
\begin{equation}
\inpd{\al}{\bt} = 0 \text{ whenever} \al\neq\bt
\label{eqn7}
\end{equation}
is satisfied, the probability of history $\al$ is given by
\begin{equation}
\Pr(\al) = \inp{\al},
\label{eqn8}
\end{equation}
assuming that $\ket{\Psi_0}$ in Eq.~\eqref{eqn6} is a normalized
state. Here $\al\neq\bt$ means that for at least one $j$ it is the case
that $\al_j\neq\bt_j$.

For a history family involving a large number of
times Eq.~\eqref{eqn7} is a fairly stringent constraint. However, for families
that involve only two times, thus one time $t_1$ following the initial time
$t_0$, it is always satisfied because of the assumed orthogonality of the
projectors at time $t_1$. In this case
\begin{equation}
\Pr(\al_1) = \mte{\Psi_0}{T(t_0,t_1)P_1^{\al_1}T(t_1,t_0)}
\label{eqn9}
\end{equation}
is simply the usual Born rule, so Eq.~\eqref{eqn8} represents a generalization
of the Born rule to histories involving three or more times, for which one has
to pay attention to Eq.~\eqref{eqn7}. Note that the probability assignment in
Eq.~\eqref{eqn8}, of which Eq.~\eqref{eqn9} is a special case, \emph{makes no
  reference to measurements}, and in this sense is distinct from what is found
in textbooks. However, by using it as a fundamental axiom one can, as for
example in Sec.~\ref{sct4}, justify all the usual textbook calculations.

What is usually found in textbooks is not Eq.~\eqref{eqn9} but the equivalent
expression
\begin{equation}
\Pr(\al_1) = \mte{\Psi(t_1)}{P_1^{\al_1}} = 
|\inpd{\Phi_1^{\al_1}}{\Psi(t_1)}|^2,
\label{eqn10}
\end{equation}
where the second equality applies when $P_1^{\al_1}$ is a rank
one projector onto the pure state $\ket{\Phi_1^{\al_1}}$. Here
\begin{equation}
\ket{\Psi(t)} := T(t,t_0)\ket{\Psi_0}
\label{eqn11}
\end{equation}
is obtained by integrating Schr\"odinger's equation starting with
$\ket{\Psi_0}$, and is often referred to as \emph{the} wave function. However,
this $\ket{\Psi(t)}$ is best viewed as a calculational tool, rather than, as
in Everett's interpretation, the actual state of the universe, or that part of
the universe that constitutes the closed quantum system under
discussion. There are two ways to see this.  One is that in general the
corresponding projector $\Psi(t) = \dya{\Psi(t)}$ does not commute with the
projector $P_1^{\al_1}$, which in the family of histories Eq.~\eqref{eqn4}
represents a particular physical property at time $t_1$. As one cannot in the
histories approach ascribe two incompatible properties to a single quantum
system at the same instant of time, $\ket{\Psi(t)}$ cannot represent a
physical property when we use the family Eq.~\eqref{eqn4}.

The second way to see that $\ket{\Psi(t_1)}$ in Eq.~\eqref{eqn10} is nothing
but a calculational tool is to calculate the same probability by another
method which makes no mention of it:
\begin{align}
\Pr(\al_1) &= |\inpd{\Phi^{\al_1}(t_0)}{\Psi_0}|^2,
\label{eqn12} \\
\noalign{\noindent where}
\ket{\Phi^{\al_1}(t)} & = T(t,t_1) \ket{\Phi_1^{\al_1}}
\label{eqn13}
\end{align}
is again obtained by integrating Schr\"odinger's equation (in the opposite
time direction) to obtain the $\ket{\Phi^{\al_1}(t_0)}$ used in
Eq.~\eqref{eqn12}. Because the same probability can be obtained using either
$\ket{\Psi(t_1)}$ or $\ket{\Phi^{\al_1}(t_0)}$, they can be regarded as
alternative calculational tools, or \emph{pre-probabilities} in the notation
of \rcite{Grff02c}, Sec.~9.4. A pre-probability is used to calculate
probabilities of physical properties, and is thus if anything less real than
the probabilities assigned to real properties. Hence ``the'' wave function
plays a very secondary role in the histories approach, quite different from
Everett or many worlds.

To be sure, there are special \emph{unitary} families of histories in which
``the'' wave function is closely associated with physical properties; an
example is given below. Before discussing it, let us note that just as there
are incompatible quantum properties, there are also incompatible histories,
and the latter must be treated in much the same way as the former. A
collection of histories associated with a suitable sample space, for example,
the $\{Y^\al\}$ defined in Eq.~\eqref{eqn4}, is referred to as a
\emph{framework} or \emph{consistent family}\footnote{The terms ``framework''
  and ``consistent family'' are used both for the sample space and the
  corresponding event algebra it generates. If the distinction is important,
  we can refer to a ``consistent sample space of histories.''} of histories
\emph{provided} the consistency conditions, Eq.~\eqref{eqn7}, for the family
under consideration, are satisfied. Two such families may be either compatible
or incompatible. They are compatible if there is a common refinement, a family
of more detailed histories which includes both families we started with, and
also satisfies the consistency conditions.  Otherwise they are incompatible,
and the single framework rule says they are not to be combined, or the
combination is meaningless (quantum theory can assign it no meaning). Once
again, the pillars of good quantum reasoning are Liberty, Equality, and
Incompatibility; see the discussion at the end of Sec.~\ref{sct2}. In
particular, Equality means that from the point of view of fundamental quantum
mechanics there is no reason to prefer one consistent family to another, there
is no law of nature that singles out ``the right family.''

How this works can be illustrated by the case of Schr\"odinger's poor cat
whose sorry history is no doubt familiar to the reader. Under unitary time
evolution the cat ends up in a state which can be described as a superposition
of dead and alive, which Schr\"odinger and his successors found perplexing. So
how does the histories approach deal with it? First, there must be a choice of
a framework, which is to say a series of possibilities at the different times
of interest. Let the initial state $\ket{\Psi_0}$ include the cat, the
apparatus, and whatever else is needed to treat this as a closed quantum
system. One possible family of histories is that in which at each time $t_j$,
$P_J^0=\dya{\Psi(t_j)}$ is a projector onto the subspace containing ``the''
wave function as defined by Eq.~\eqref{eqn11}; $P_j^1 = I - P_j^0$ is its
negation. This family is unitary in the notation in \rcite{Grff02c}, Sec.~8.7;
it is consistent, and the physicist who wants to use it is at Liberty to do
so.  But the historian will remind him that referring to ``Schr\"odinger cat''
states at later times constitutes something of a misnomer because the
corresponding projector $P_j^0$ will (at least in general) not commute with
projectors for physical properties of anything one would want to call a
cat. If the physicist --- says the historian --- wants to discuss what is
happening to something that can properly be called a cat, then it is necessary
to adopt a different framework, one containing projectors referring to states
in which the cat is dead or alive as mutually exclusive possibilities.  Such
frameworks exist, are just as fundamental (Equality) as the unitary framework,
and of much greater utility to the physicist interested in biology or some
other aspect of the ordinary macroscopic world.

\section{Measurements}
\label{sct4}

One of the unfortunate features of contemporary quantum mechanics textbooks is
their inability to make much sense of experiments of the sort that are done
all the time in the laboratory by competent experimental physicists. Such
measurements are typically designed to reveal properties possessed by systems
\emph{before} the measurement takes place. What does this electrical pulse
mean? It means that the proton knocked out of the target by the high energy
electron has ionized the gas near a pair of wires in the detector\ldots. By
the time they give talks reporting their results experimental physicists seem
to have forgotten (fortunately) what they learned from textbooks about
measurements magically producing reality out of nothing.  When the quantum
analysis is done properly using histories, one finds that the experimentalists
are correct. Once one allows them Liberty their discussions make perfectly good
sense.

A simple schematic example of an idealized measuring process can be used to
illustrate the essential points. We assume the Hilbert spaces of the system to
be measured and the measuring apparatus are $\HM_s$ and $\HM_M$, and that at
the earliest time, $t_0$, the two are in a product state
$\ket{s}\ot\ket{M_0}$. The unitary time development operator $T(t',t)$ is the
identity between $t_0$ and $t_1$, but in the interval from $t_1$ to $t_2$ the
system interacts with the apparatus as indicated here, with time steps in the
order $t_0\ra t_1\ra t_2$:
\begin{equation}
\ket{\Psi^j} = \ket{s^j}\ot\ket{M_0} \ra \ket{s^j}\ot\ket{M_0} \ra \ket{\bar s^j}\ot\ket{M^j}.
\label{eqn14}
\end{equation}
The $\{\ket{s^j}\}$ form an orthonormal basis of $\HM_s$, assumed to be a
Hilbert space of finite dimension,\footnote{We think of $\HM_s$ as
  corresponding to internal states, say spin states, of a particle, and that
  its center of mass motion toward a detector is taken care of by using a
  unitary time development operator $T(t',t)$ with two arguments, rather than
  supposing it depends only on the time difference $t'-t$.}  the
$\{\ket{M^j}\}$ are orthogonal to each other, and the $\{\ket{\bar s^j}\}$ are
normalized states in $\HM_s$, but not necessarily orthogonal. We do \emph{not}
have to assume that $\ket{\bar s^j}=\ket{s^j}$.

Assume an initial state
\begin{equation}
\ket{\Psi_0} = \ket{s}\ot\ket{M_0} = \Blp \sum_j c_j\ket{s^j}\Brp\ot\ket{M_0}
\label{eqn15}
\end{equation}
at $t_0$, and a family of histories [see Eq.~\eqref{eqn4}],
\begin{equation}
\Psi_0\od \{s^j\} \od \{M^k\}.
\label{eqn16}
\end{equation}
As in Eq.~\eqref{eqn1} the letters denote projectors: $s^j=\dya{s^j}$,
etc. Each history begins with the initial state $\ket{\Psi_0}$ at $t_0$. At
time $t_1$ the particle is in one of the states $\ket{s^j}$, and at $t_2$ the
apparatus pointer is in the position indicated by the projector $M^k$. Notice
that these histories contain no reference to the measuring apparatus at time
$t_1$; think of $s^j$ as the same as $s^j\ot I_M$. Nor is there any reference
to the state of the particle at time $t_2$. Given the dynamics in
Eq.~\eqref{eqn14} it is straightforward to show that the consistency
conditions, Eq.~\eqref{eqn7}, are satisfied and the probabilities are, in an
obvious notation where subscripts denote the time,
\begin{equation}
\Pr(s^j_1,M^k_2) = |c_j|^2 \dl_{jk}.
\label{eqn17}
\end{equation}
From these one can conclude by summing over $j$ that the probability that the
pointer is in state $M^k$ at $t_2$ is equal to $|c_k|^2$, the same result as
obtained by the usual textbook rules. But Eq.~\eqref{eqn17} also yields a
conditional probability:
\begin{equation}
\Pr(s^j_1\vb M^k_2) = \dl_{jk}.
\label{eqn18}
\end{equation}
That is, if the pointer is in position $M^k$ at time $t_2$, we can be sure
that the system was in the state $s^k$ at the earlier time $t_1$. 
The apparatus was doing what it was designed to do.

Notice that this analysis makes no assumption about the nature of the
$\ket{\bar s^j}$ states that appear at time $t_2$ in the dynamics given in
Eq.~\eqref{eqn14}. They are irrelevant if one regards, as is typically the
case, the purpose of the measurement to be that of determining some property
of the measured system \emph{before} interaction with the apparatus, which may
well have altered the property in question, or even destroyed the system (as,
for example, when a photon is absorbed). However, we can also consider the
special case of a nondestructive measurement for which $\ket{\bar s^j} =
\ket{s^j}$ for every $j$. In this case we may replace the projectors at the
final time $t_2$ in Eq.~\eqref{eqn16} with $\{s^l\ot M^k\}$, and after a short
calculation derive the result
\begin{equation}
\Pr(s_1^j,M_2^k,s_2^l) = |c_j|^2 \dl_{jk}\dl_{kl}.
\label{eqn19}
\end{equation}
Thus given that the pointer is in the state $M^k$ at $t_2$ we can deduce
the state of the system $s$ both before and after the measurement.

It is important to observe that the results given here, which agree with the
results of textbook calculations and the belief of experimentalists that they
can design good measurements, are \emph{not} based on a ``measurement
postulate'' of the sort made by von Neumann and later textbook
writers. They are a consequence of Eq.~\eqref{eqn8} in a situation in which
the consistency conditions, Eq.~\eqref{eqn7}, apply. These are fundamental
postulates that apply to any closed quantum system, which is to say to any
situation in which the dynamics is determined by solving Schr\"odinger's
equation, not just cases in which a measurement is taking place.

But it is also worth noting the sense in which both von Neumann and the
textbooks are to some degree correct. From the histories perspective they can
be understood as using a family of histories of the form, focusing for
simplicity on the nondestructive case,
\begin{equation}
\Psi_0\od \{\Psi_1, I-\Psi_1\} \od \{s^l\ot M^k\},
\label{eqn20}
\end{equation}
where $\Psi_1$ is the projector onto the unitarily-evolved state $\ket{\Psi_1}
= T(t_1,t_0)\ket{\Psi_0}$. It is easily checked that this family is
consistent, and physicists are at Liberty to employ it. It provides just as
good a description of nature as Eq.~\eqref{eqn16}. However, the two families
are incompatible if at least two of the $c_j$ in Eq.~\eqref{eqn15} are
nonzero; in particular $\Psi_1$ will (in general) not commute with $s^j$. This fact means that using the family Eq.~\eqref{eqn20} precludes
saying anything meaningful about the properties of $s$ before the measurement
takes place. So the reticence of textbook authors has some justification:
their approach is unable to connect measurement outcomes with earlier
properties, because to do so a framework is needed in which the relevant
properties make sense. The problem is that the very possibility of alternative
descriptions, the principles of Liberty and Equality, are absent from textbook
treatments.

To summarize, in order to discuss measurements \emph{as measurements} in the
typical sense in which experimental physicists think of them, it is necessary
to introduce frameworks or families in which \emph{both} the measurement
outcome \emph{and} the relevant properties of the system before the
measurement takes place are made part of a consistent family. Textbooks in
effect adopt a framework in which outcomes make sense, because otherwise their
discussions would never get off the ground: simple use of unitary time
development leads to macroscopic quantum superpositions and the measurement
problem. But then there is no good reason why, having taken one Liberty, a
second, namely frameworks in which properties make sense before the
measurement interaction, should be denied. And this second Liberty is used all
the time by the experimentalists; it is past time for theoreticians and
textbook writers to catch up.  There is more to be said about wave function
collapse, but we defer this to an appropriate point in the following
discussion.

\section{EPR Correlations}
\label{sct5}

\subsection{Entangled states}
\label{sbct5.1}

Let us now turn to the EPR problem\cite{EnPR35} and its connection with
locality, focusing on the standard example due to Bohm\cite{Bhm51s} of two
spin-half particles $a$ and $b$ in the state described by the singlet wave
function
\begin{equation}
\ket{\psi_0} = \Blp\ket{z_a^+z_b^-} - \ket{z_a^-z_b^+}\Brp/\st,
\label{eqn21}
\end{equation}
where $\ket{z_a^+z_b^-}=\ket{z^+}_a\ot\ket{z^-}_b$ is an eigenstate of both
$S_{az}$ for the $a$ particle, eigenvalue $+1/2$, and of $S_{bz}$ for the $b$
particle, eigenvalue $-1/2$; the state $\ket{z_a^-z_b^+}$ has a similar
interpretation with eigenvalues of the opposite sign. By contrast, the
superposition represented by $\ket{\psi_0}$ is not an eigenstate of either
$S_{az}$ or $S_{bz}$. Indeed, the corresponding projector $\psi_0=
\dya{\psi_0}$ is incompatible with the property $S_{aw}=+1/2$ of particle $a$,
where $w$ is any direction in space, and likewise with $S_{bw}=+1/2$. Thus it
makes no sense to say that a quantum system which possesses the property
$\psi_0$ has any nontrivial property corresponding to particle $a$ or particle
$b$, that is, to some subspace of the corresponding Hilbert space. (The
trivial properties are the identity and zero projectors, which are,
respectively, always true and always false.)

Before going further it is worth remarking that there is no notion of
nonlocality intrinsic to the singlet state $\ket{\psi_0}$ as such. After all,
it is the spin state of a hydrogen atom or of positronium in its ground state,
where one does not usually think of the particles as in distinct locations.
Entanglement as such is a quantum concept distinct from any notion of
nonlocality. To tie $\ket{\psi_0}$ to nonlocality, one has to to suppose that
the two particles are in different locations. Of course, just this sort of
thing is frequently achieved in the laboratory nowadays, and, when it is, such
states can with some justification be called \emph{nonlocal}.

Nonlocality of this sort is, however,  not what Bell and his successors have
had in mind when speaking of nonlocality in quantum mechanics, for one does
not need carefully constructed inequalities to demonstrate that entangled
states are present in the quantum Hilbert space. The issue is not one of
\emph{static} nonlocality, but instead whether quantum theory allows or
demands certain nonlocal \emph{dynamical} effects. As Maudlin puts it, p.~123
of \rcite{Mdln10}, the locality assumption which he thinks quantum mechanics
violates is that ``the measurement on one particle does not change the state
of the other [particle].''

\subsection{Dynamics}
\label{sbct5.2}

We begin our study by considering the dynamics of the spin degrees of
freedom of particles $a$ and $b$, making the usual assumption that the
particles are not interacting with each other and no magnetic fields are
present, and hence the time development operator
\begin{equation}
T(t',t)=T_a(t',t)\ot T_b(t',t) = I_a\ot I_b
\label{eqn22}
\end{equation}
is trivial. Later we will explore the effects of measurements, but first
let us examine what happens in their absence.

Consider the family
of histories based on three times $t_0<t_1<t_2$,
\begin{equation}
\psi_0\;\od \;\{z_a^+,z_a^-\}\ot\{z_b^+,z_b^-\} 
\;\od\;\{z_a^+,z_a^-\}\ot\{z_b^+,z_b^-\},
\label{eqn23}
\end{equation}
which is to be interpreted as follows. All the
histories begin at time $t_0$ with the initial state $\psi_0=\dya{\psi_0}$. At
time $t_1$ there are four possible properties or ``events'': $z_a^+z_b^+
=z_a^+\ot z_b^+$, which means $S_{az}=+1/2$, $S_{bz}=+1/2$, $z_a^+z_b^-$,
$z_a^-z_b^+$, and $z_a^-z_b^-$. The same four possibilities occur at the later
time $t_2$. Consistency can be checked and probabilities assigned to the
$4\tm 4=16$ histories as explained in Sec.~\ref{sct3}. Only two histories have
nonzero probabilities:
\begin{equation}
\Pr(z_{a1}^+,z_{b1}^-;z_{a2}^+,z_{b2}^-) = 
\Pr(z_{a1}^-,z_{b1}^+;z_{a2}^-,z_{b2}^+) = 1/2.
\label{eqn24}
\end{equation}
Here semicolons rather than commas are used to separate events at different
times, indicated by subscripts, to improve legibility. Both semicolons and
commas should be read as ``\AND.''

Summing over the values at $t_1$ yields the marginal probabilities
\begin{equation}
\Pr(z_{a2}^+,z_{b2}^-) = \Pr(z_{a2}^-,z_{b2}^+) = 1/2
\label{eqn25}
\end{equation}
at time $t_2$.  These can be obtained from $\ket{\psi_0}$ by use of the Born
rule (see Sec.~\ref{sct3}). There is no need to refer to
measurements. Properly built apparatus, as discussed in Sec.~\ref{sbct5.3}
below, reveals what is there before the measurement takes place. But
Eq.~\eqref{eqn24} tells us more than Eq.~\eqref{eqn25}. For example, it
implies that the values of $S_{az}$ are identical at times $t_1$ and $t_2$, a
result that is physically reasonable but not implied by the Born rule.

Liberty allows many other families besides Eq.~\eqref{eqn23}. If $z$ is
replaced everywhere in Eq.~\eqref{eqn23} by $w$, where $w$ could be $x$ or $y$
or any other direction in space, it follows from the spherical symmetry of
$\ket{\psi_0}$ that the family
\begin{equation}
\psi_0\;\od \;\{w_a^+,w_a^-\}\ot\{w_b^+,w_b^-\} \;\od\;\{w_a^+,w_a^-\}\ot\{w_b^+,w_b^-\},
\label{eqn26}
\end{equation}
is consistent, yielding the same probabilities in Eq.~\eqref{eqn24} and
marginals in Eq.~\eqref{eqn25} if $w$ is replaced by $z$ in these
expressions. Perhaps of greater interest is the family
\begin{equation}
\psi_0\;\od \;\{z_a^+,z_a^-\}\ot\{w_b^+,w_b^-\} 
\;\od\;\{z_a^+,z_a^-\}\ot\{w_b^+,w_b^-\},
\label{eqn27}
\end{equation}
with $S_z$ for particle $a$, but $S_w$, with $w$ an arbitrary but specific
direction, for particle $b$. The family is consistent, but now four
histories have nonzero probability:
\begin{align}
\Pr(z_{a1}^+,w_{b1}^-;z_{a2}^+,w_{b2}^-) &= 
\Pr(z_{a1}^-,w_{b1}^+;z_{a2}^-,w_{b2}^+)
\notag\\
&= (1/2)\cos^2(\th/2),
\notag\\
\Pr(z_{a1}^+,w_{b1}^+;z_{a2}^+,w_{b2}^+) &=
\Pr(z_{a1}^-,w_{b1}^-;z_{a2}^-,w_{b2}^-) 
\notag\\
&= (1/2)\sin^2(\th/2).
\label{eqn28}
\end{align}
Note that whatever value $S_{bw}$ has at time $t_1$, it has exactly the same
value (with probability 1) at time $t_2$, just as one would have expected for
a particle in the absence of any interaction with the rest of the world.
Again the marginals at time $t_2$, $\Pr(z_{a2}^+,w_{b2}^-)$ and the like, have
values given by the Born rule. And note that the three consistent families in
Eqs.~\eqref{eqn23}, \eqref{eqn26}, and \eqref{eqn27} are all mutually
\emph{incompatible}, except for the special case in which $w$ is equal to $z$
or to $-z$; there is no meaningful way to combine the corresponding
probabilities.

\subsection{Measurements}
\label{sbct5.3}

Consider a measurement of $S_z$ on particle $a$. In any fundamental
quantum analysis the measuring apparatus must also be included as part of
the total system, and described in fully quantum mechanical terms. Thus we use a
Hilbert space $\HM_a\ot\HM_M\ot\HM_b$, and assume that for $t$ and $t'$ in
the range of interest the total unitary time development operator
\begin{equation}
T(t',t) = T_{aM}(t',t)\ot T_b(t',t)
\label{eqn29}
\end{equation}
factors into a piece $T_{aM}$ in which particle $a$ interacts with the
apparatus, and a piece $T_b(t',t)=I_b$: nothing is happening to particle $b$.
Further assume that the action of $T_{aM}(t',t)$ is given for the range
$t_0\ra t_1\ra t_2\ra t_3$ by
\begin{align}
\ket{z_a^+}\ot\ket{M_0}&\ra \ket{z_a^+}\ot\ket{M_0}
\notag\\
&\ra \ket{z_a^+}\ot\ket{M_0}\ra \ket{\bar z_a^+}\ot\ket{M^+},
\notag\\
\ket{z_a^-}\ot\ket{M_0}&\ra \ket{z_a^-}\ot\ket{M_0}
\notag\\
&\ra \ket{z_a^-}\ot\ket{M_0} \ra\ket{\bar z_a^-}\ot\ket{M^-},
\label{eqn30}
\end{align}
where the arrows indicate how the ket to the left evolves unitarily to the ket
on the right during the corresponding time interval. The final states
$\ket{\bar z_a^+}$ and $\ket{\bar z_a^-}$ at $t_3$ are arbitrary and could be
omitted from the discussion (see remarks in Sec.~\ref{sct3}); the reader
who wants to remove the bars is welcome to do so.

Consider the family of histories at times $t_0<t_1<t_2<t_3$
\begin{align}
\Psi_0\;&\od\; \{ z_a^+,z_a^-\}\ot \{z_b^+, z_b^-\}\;\od\; \{ z_a^+,z_a^-\}\ot \{z_b^+, z_b^-\}\;
\notag\\
& \quad \od\; \{M^+,M^-\}\ot\{z_b^+, z_b^-\},
\label{eqn31}
\end{align}
where $\Psi_0$ is the projector on the initial state $\ket{\Psi_0} =
\ket{\psi_0}\ot\ket{M_0}$ of the system at $t_0$: a singlet spin state
together with an apparatus ready to measure $S_{az}$. As in the family in
Eq.~\eqref{eqn24}, there are just two histories with nonzero probabilities:
\begin{align}
\Pr(z_{a1}^+,z_{b1}^-;z_{a2}^+,z_{b2}^-;M_3^+,z_{b3}^-) &= \notag\\
\Pr(z_{a1}^-,z_{b1}^+;z_{a2}^-,z_{b2}^+;M_3^-,z_{b3}^+) &= 1/2.
\label{eqn32}
\end{align}
From these we can calculate marginal and conditional probabilities using the
usual rules of ordinary probability theory. In particular,
\begin{equation}
\Pr(z_{aj}^+\vb M_3^+) = \Pr(z_{bk}^-\vb M_3^+) = 1,
\label{eqn33}
\end{equation}
with $j=1$ or 2 and $k=1$, 2, or 3. That is, from the measurement outcome $M^+$
at time $t_3$ one can infer that particle $a$ at earlier times (but later than
$t_0$) had $S_z=+1/2$, while particle $b$ had $S_z=-1/2$ at times $t_1$ and
$t_2$, and continued to possess this value at time $t_3$. All these results
are reasonable in light of the earlier analysis of
Eq.~\eqref{eqn23} and the probabilities in Eq.~\eqref{eqn24}. Properly
designed measurements reveal what is there to be measured; they do not somehow
create reality out of a vacuum.

All these results are inaccessible using the calculational rules of standard
textbooks, except for the result $\Pr(z_{b3}^-\vb M_3^+)=1$. One of the
standard rules, which students find rather odd and ad hoc, states that if the
outcome of a measurement is, say, $M^+$ at $t_3$, then one should ``collapse''
(first arrow) and renormalize (second arrow) the singlet state wave function:
\begin{equation}
\ket{\psi_0} \ra z_a^+\ket{\psi_0} = (1/\st)\ket{z_a^+ z_b^-} \ra
\ket{\psi^+} :=\ket{z_a^+ z_b^-},
\label{eqn34}
\end{equation}
and then calculate
\begin{equation}
\Pr(z_{b3}^-\vb M_3^+) = \mte{\psi^+}{z_{b}^-} = 1.
\label{eqn35}
\end{equation}
This is a perfectly good calculational rule, and it can be justified, at least
as used in the present context, on the basis of fundamental quantum
principles, though we shall not take the time to do so here.
\footnote{Strictly speaking, Eq.~\eqref{eqn34} should be applied only in cases
  in which the bars are absent in the final $a$ states in the dynamics in
  Eq.~\eqref{eqn30}. However, if we are only using the collapsed wave function
  to calculate properties of particle $b$ this makes no difference.} Like
most calculational rules it allows students to obtain the right answer without
having to think about what they are doing, which is in effect calculating a
conditional probability. The disadvantage is that they may mistakenly come to
believe that wave function collapse is a physical process rather than a
calculational tool. That there is no nonlocal physical effect associated with
it can be seen by noting that the same ``collapse'' approach also yields the
right answer, Eq.~\eqref{eqn33} for $S_{bz}$ at the earlier times $t_1$ and
$t_2$ \emph{before} the measurement of $S_{az}$ occurs. And for those
interested in relativistic quantum mechanics we remark that at time $t_1$
particle $b$ can be in the backward light cone of the spacetime region, with
$t$ somewhere between $t_2$ and $t_3$, in which the measuring device interacts
with particle $a$. Backward as well as superluminal causation readily emerges
from quantum theory when calculational methods are confused with physical
causes.

Rather than Eq.~\eqref{eqn31} one can analyze the family
\begin{align}
&\Psi_0\;\od\; \{ z_a^+,z_a^-\}\ot \{w_b^+, w_b^-\}\;
\od\;\hphantom{\{ z_a^+,z_a^-\}}
\notag\\
& \quad \{ z_a^+,z_a^-\}\ot \{w_b^+, w_b^-\}\;
\od\; \{M^+,M^-\}\ot
\{w_b^+, w_b^-\},
\label{eqn36}
\end{align}
where the focus is now on $S_{bw}$ in place of $S_{bz}$, with $w$ an
arbitrary direction in space making an angle $\th$ with the positive $z$ axis,
as in Eq.~\eqref{eqn27}.
The resulting conditional probabilities include
\begin{align}
\Pr(z_{aj}^+\vb M_3^+) &= 1,\quad \Pr(w_{bk}^+\vb M_3^+) = \sin^2(\th/2),
\label{eqn37}\\
\Pr(w_{bk}^+\vb w_{b1}^+) &= \Pr(w_{bk}^-\vb w_{b1}^-) =1,
\label{eqn38}
\end{align}
for $j=1,2$ and $k=1,2,3$. These make perfectly good physical sense. In
particular Eq.~\eqref{eqn38} tells us that nothing at all is happening
to $S_{bw}$ at any time during the interval between $t_2$ and $t_3$ when
particle $a$ is interacting with the measuring apparatus. Local measurements
properly analyzed have no nonlocal effects.

One can also introduce a second measuring apparatus that measures $S_{bw}$ for
particle $b$, with $T_{bN}(t',t)$ for each time interval in $t_0\ra t_1\ra
t_2\ra t_3$ equal to the identity, except $t_2\ra t_3$ given
by [compare Eq.~\eqref{eqn30}]
\begin{equation}
\ket{w_b^+}\ot\ket{N_0}\ra \ket{\bar w_b^+}\ot\ket{N^+},\quad
\ket{w_b^-}\ot\ket{N_0} \ra\ket{\bar w_b^-}\ot\ket{N^-}.
\label{eqn39}
\end{equation}
Readers should have no difficulty checking the
consistency and working out the probabilities associated with the family
\begin{align}
&\Psi_0\;\od\; \{ z_a^+,z_a^-\}\ot \{w_b^+, w_b^-\}\;\od\;
\hphantom{\{ z_a^+,z_a^-\}}
\notag\\
&\{ z_a^+,z_a^-\}\ot \{w_b^+, w_b^-\}\;\od\; \{M^+,M^-\}\ot \{N^+,N^-\},
\label{eqn40}
\end{align}
where $\Psi_0$ is now the projector on $\ket{\Psi_0} =
\ket{\psi_0}\ot\ket{M_0}\ot\ket{N_0}$, and deriving conditional probabilities
such as
\begin{align}
\Pr(z_{aj}^+\vb M_3^+) &= \Pr(z_{aj}^-\vb M_3^-) =1,
\label{eqn41}\\
\Pr(w_{bj}^+\vb N_3^+) &= \Pr(w_{bj}^-\vb N_3^-) = 1,
\label{eqn42}\\
\Pr(N_3^+\vb M_3^+) &= \Pr(N_3^-\vb M_3^-) = \sin^2(\th/2),
\label{eqn43}
\end{align}
with $j=1,2$. Textbook rules yield Eq.~\eqref{eqn43}, minus any insights
obtained by relating the measurement outcomes to the particle properties they
were designed to measure.

\section{Discussion}
\label{sct6}

\subsection{Locality}
\label{sbct6.1}

What has been shown is that ``The Predictions,'' as Maudlin\cite{Mdln10} calls
them, of quantum mechanics, in particular Eq.~\eqref{eqn43}, are produced by a
theory in which local measurements have no nonlocal effects: the measurement
of $S_{az}$ that occurs between $t_2$ and $t_3$ has no effect on $S_{bw}$, see
Eq.~\eqref{eqn38}. This demonstrates that there is something amiss with the
version of Bell's argument found in \rcite{Mdln10}, and because it is compact,
we can hope to locate where the reasoning departs from the consistent
principles of quantum mechanics discussed in Secs.~\ref{sct2}--\ref{sct4}.

My summary of Maudlin's argument in Sec.~II~C of \rcite{Mdln10}, with the
notation changed to make it consistent with that in Sec.~\ref{sct5}, is as
follows:

\begin{description}

\item[M1] Measurements of a given spin component on particles $a$ and $b$ will
always yield opposite outcomes: $+1/2$ for one particle is perfectly
correlated with $-1/2$ for the other.

\item[M2] Assume locality: A measurement carried out on particle $a$ cannot
affect the physical state of particle $b$.

\item[M3] This means that particle $b$ must already have been disposed to
  yield the opposite result even before particle $a$ was measured.

\item[M4] Particle $b$ must have been so disposed even when the total quantum
  mechanical state of the system was the singlet state $\ket{\psi_0}$, and at
  all times since its creation.

\item[M5] Therefore the complete physical description of particle $b$ must
determine how it is disposed to yield a particular outcome for each
possible spin measurement, because M1 holds for any spin component.

\item[M6] Hence a local theory must not only be deterministic but also a
hidden variable theory (in the sense of M5).

\item[M7] Any local theory that predicts the EPR correlations must also
respect certain constraints on the correlations it predicts (Bell's
inequality).

\item[M8] Quantum mechanics violates these constraints and thus the locality
assumption in M2.

\end{description}

Let us compare these assertions with the analysis in
Sec.~\ref{sct5}. Evidently M1 corresponds to the anticorrelation expressed in
Eq.~\eqref{eqn25} in terms of the particles themselves, or in
Eq.~\eqref{eqn43} for measurement outcomes, assuming $\th=0$, that is,
measurements of the same spin component. Note that although these formulas
were obtained for $S_{az}$ and $S_{bz}$, the results are equally valid if $z$
is replaced, for both particles, with an arbitrary direction
$w$. \footnote{Naturally, one has to use a different piece of apparatus to
  measure different components of the spin angular momentum.} As for M2,
dynamical locality is ensured by the fact that the time development operator
is a tensor product of the form $T_a\ot T_b$ for the particles alone, or
$T_{aM}\ot T_{bN}$ when measuring apparatuses are included. Furthermore,
direct calculation, see Eq.~\eqref{eqn38} and the comments following it, shows
that the spin of particle $b$ is unaffected by making a measurement on
particle $a$. And M3, the disposition of the $b$ particle to yield the
opposite result even before the $a$ particle was measured, is evident from the
fact that Eqs.~\eqref{eqn33} or \eqref{eqn37} hold for all applicable values
of $t_j$ and $t_k$: $t_1$ and $t_2$ before the measurement of particle $a$ and
$t_3$ after the measurement. This disposition resides in the simple fact that
the particle actually had at the earlier time the property which a measurement
would later reveal. Thus far everything is fine.

But with M4 we run into difficulties. If by ``state of the system'' we are to
understand $\ket{\psi_0}$ as the physical property possessed by the system,
then, as noted in Sec.~\ref{sct2}, there is no meaningful way to ascribe a
value to any component of the spin angular momentum of particle $b$ at this
time, because the two projectors do not commute. Ignoring this is to ignore a
crucial difference, one might say the crucial difference, between the quantum
and the classical world: quantum incompatibility. One way out of the
difficulty is to consider $\ket{\psi_0}$ a pre-probability rather than a
property; see the discussion following Eq.~\eqref{eqn13}, and in
\rcite{Grff02c}, Sec.~9.4. But it is not clear that a reference to
$\ket{\psi_0}$ is really essential for the later stages of Maudlin's argument,
so perhaps we can replace M4 by a modified form consistent with the framework
Eq.~\eqref{eqn31}:
\begin{description}
\item[M4$'$] Particle $b$ was disposed to yield the opposite result of the measurement of particle $a$ at all times beginning shortly after the interaction of the two particles produced the singlet state.
\end{description}

However, at M5 we arrive at a significant divergence from the principle of
Hilbert space quantum mechanics that states that any physical description of a
particle at a single time must correspond to some subspace of its Hilbert
space. Neither $\ket{\psi_0}$ nor any subspace of $\HM_b$ for particle $b$ can
be interpreted as indicating how particle $b$ is disposed to yield a
\emph{particular} outcome for \emph{each} possible spin measurement. This is
in contrast to a ``classical'' spinning object, such as a golf ball, where one
can easily imagine a spin state which can be used to predict, at least with
very high precision, the outcome of a measurement of any component of its spin
angular momentum. And there is no obvious way of rewording M5 to evade this
difficulty.

What has led to this problem? Let us return to M1 and ask in what sense it
holds, as asserted in M5, for \emph{any} spin component. Given a
\emph{particular} $w$, we can choose a consistent family of the form in
Eq.~\eqref{eqn26}, add a measurement apparatus if desired, and come to the
conclusion that $S_{aw}=-S_{bw}$. It is therefore natural to assume, and it
would be true in a classical world, that $S_{aw}=-S_{bw}$ is
\emph{simultaneously} true for \emph{every} $w$. But different $w$ lead to
different, incompatible frameworks, and therefore this last conclusion
violates the single framework rule of consistent quantum reasoning.  Nothing
similar ever arises in pre-quantum physics, so it is easy to make this kind of
mistake by employing classical reasoning in a domain where it does not work,
namely, Hilbert-space quantum mechanics.

The foregoing analysis indicates what seems to be the critical misstep in the
presentation of Bell's argument in \rcite{Mdln10}: classical reasoning applied
to the description of microscopic quantum particles in a way that violates the
principles of Hilbert space quantum mechanics, as expressed by the single
framework rule. As noted in Sec.~\ref{sct2}, Birkhoff and von Neumann were
aware that Hilbert space quantum mechanics is incompatible with classical
modes of reasoning. Ignoring this insight does not make the problem go away,
and numerous quantum paradoxes, along with the claim that the quantum world is
pervaded with nonlocal influences that violate relativity theory, can be seen
to have their roots in a failure to deal seriously with the logical problem
posed by a Hilbert space description. But the connection of all this with
counterfactual arguments deserves some additional discussion.

\subsection{Counterfactuals}
\label{sbct6.2}

Blaylock\cite{Blyl10} claims that a key component in Bell's inequality, or at
least an argument leading to it, is counterfactual definiteness: the
assumption that a measurement that was not performed had a single definite
result. He concludes that therefore the violation of Bell's inequality by
quantum mechanics does not by itself imply nonlocality. Does our
analysis throw some light on this matter?

It is helpful to explore the issue of counterfactuals in quantum mechanics
starting with a simpler example than the one considered in Sec.~\ref{sct5},
namely a measurement of a component of spin angular momentum in a situation in
which the apparatus can be adjusted to measure either $S_z$ or $S_x$. For
example, for a particle traveling along the $y$ axis one could imagine
rotating the Stern-Gerlach magnet so that the field gradient is parallel to
$z$ or to $x$. Naturally, only one component can be measured in a given run;
suppose that in a particular case it was $S_z$. Then the following
counterfactual question seems sensible: What \emph{would have} been the result
\emph{if} the apparatus had been set up to measure $S_x$ instead? And we can
ask: is the answer to this question different for the (real) quantum world
than it would be for a (hypothetical) classical world?  It is possible to set
up a model in which the measurement axis is determined by the outcome of
``flipping a quantum coin,'' and address the situation inside the total closed
system (coin plus apparatus plus particle being measured) using the
appropriate quantum analysis to calculate the probabilities for a single
consistent family; see \rcite{Grff02c}, Chap.~19.  There is no problem finding
a decomposition of the identity which includes the apparatus pointer positions
for both outcomes of the quantum coin flip. But to describe this as an
authentic measurement in which the pointer position is correlated with a state
before the measurement we need to include the corresponding particle property
at a time before the measurement took place. In the case of the $S_z$
measurement the corresponding particle projectors will be $z^+$ and $z^-$,
whereas for the counterfactual $S_x$ measurements we would like to use $x^+$
and $x^-$.  However, the $z$ projectors do not commute with the $x$
projectors, so they cannot both be placed in the same consistent family. Note
that the issue raised here has no direct connection with locality; we are
considering only a single measurement on a single particle at a single
location.

Valid counterfactual reasoning in the quantum domain ought to follow the
same standards as ordinary reasoning, which is to say it must be restricted to
a single consistent family. Thus the incompatibility at the microscopic level
discussed in the previous paragraph can render a quantum counterfactual
argument invalid even if the corresponding classical argument is correct, or
at least plausible. There is never a problem if we limit the discussion to
measurement outcomes and exclude all talk, even implicit, of particle
properties preceding the measurement. However, Bell inequality derivations
always make some assumption about what goes on in the world prior to the
measurements themselves, and Maudlin's argument for nonlocality is filled with
references to particle properties, so in either case counterfactual parts of
the argument, if present, could well be in conflict with quantum
theory. Even in the classical domain, counterfactual arguments can make an
implicit reference to a preceding state of affairs, which may make it
difficult to analyze their structure; see \rcite{Grff02c}, Sec.~19.3. So is
is plausible that something like that will also be present when these
arguments are applied to quantum situations.

In fact it is possible to arrive at Maudlin's M5, which as we pointed is in
conflict with quantum theory as consistently interpreted using histories, by
employing the following counterfactual argument: 

We know that if $S_z$ is measured for particle $a$ and the outcome is
$S_{az}=+1/2$, then $S_{bz}$ was $-1/2$ even before a measurement which
confirmed that it had this value. If on the other hand $S_{aw}$, with $w$ some
other direction in space, had been measured with, say, the outcome $+1/2$,
then by the same line of reasoning $S_{bw}$ would have had the value $-1/2$,
which would have been confirmed by a later measurement had it measured
$S_{bw}$. So the $b$ particle must have had a definite value of $S_{bw}$ for
every possible $w$ before any measurement took place, namely, the opposite of
the outcome of the $S_{aw}$ measurement which did not but could have taken
place.

This argument is of the counterfactual type because it starts with the
assumption that $S_{az}$ was measured in the actual world and $S_{aw}$ in the
imaginary or counterfactual world. In it one sees a connection with what
Blaylock considers suspicious about derivations of Bell's inequality:
counterfactual definiteness, the assumption that unperformed measurements have
definite outcomes. The historian will point to a failure to follow the single
framework rule as the most likely error on the route from seemingly reasonable
assumptions to a conclusion that is obviously wrong, at least if Hilbert space
quantum mechanics is accepted as the physicist's fundamental description of
the world; see the discussion of M5 in Sec.~\ref{sbct6.1} above. One cannot be
sure that Maudlin was using this kind of argument to arrive at what I have
called M5, although the following quotation from \rcite{Mdln10}, p.~123, the
second sentence of which is to a degree counterfactual (and contains a whiff
of free choice), suggests it may not have been all that far from his thinking:
\begin{quote}
  ``The \emph{only} way to give a local physical account of the EPR
  correlations is for each of the particles to be initially disposed to yield
  a particular outcome for each possible spin measurement. For if either
  particle is not so disposed and if we happen to measure the spin in the
  relevant direction (as we might), then there could be no guarantee that the
  outcomes of the two measurements will be anticorrelated.''
\end{quote}

In summary it seems plausible that flawed (from the quantum perspective)
counterfactual reasoning provides at least one possible route for deriving
Bell's inequality, with its conclusions inapplicable to the real (quantum)
world, and in this respect the analysis given in this paper supports
Blaylock's conclusions. For more on the subject of the correct use of
counterfactuals in the quantum domain, see \rcite{Grff02c}, Chap.~19.

\section{\ Conclusion}
\label{sct7}

What we have shown by means of a counterexample is that quantum violations of
Bell's inequality are perfectly consistent with quantum mechanics being a
local theory, in the sense that measurements near one point in space do not
immediately affect what goes on elsewhere. So the claim by Maudlin\cite{Mdln10}
that no local theory can reproduce the predictions of
quantum mechanics is incorrect, at least for the situation under discussion:
two spin-half particles initially in a singlet state.  Much more can be said
about quantum locality, and \rcite{Grff11} gives a more extensive treatment of
the topic using the histories formulation in a consistent manner to sort out
the issues. There the suggestion is made that Bell's inequality is best
thought of as being appropriate to the domain of classical physics but not
quantum physics: it has a perfectly good derivation, at least for all
practical purposes, in the case of golf balls, and one can see explicitly how
this derivation breaks down as the total angular momentum quantum number
decreases from around $10^{30}$ to $1/2$. Thus there is nothing wrong
with Maudlin's reasoning, except that it is does not apply to the quantum
world. Here new rules of reasoning are needed, as Birkhoff and von
Neumann\cite{BrvN36} realized, even though their particular proposal has not
turned out to be a fruitful approach for understanding quantum mechanics.

The extent to which counterfactual reasoning of a sort inconsistent with
quantum principles is a necessary part of derivations of Bell's inequality is
less clear, though this is certainly one route for getting there, as
Blaylock\cite{Blyl10} pointed out. In particular, it is plausible that
Maudlin's version of the Bell argument makes use of some form of
counterfactual reasoning, and in this sense the present article supports the
conclusion of Blaylock that counterfactual reasoning, rather than an
assumption of locality, is why derivations of Bell's inequality lead to
conclusions inconsistent with quantum theory and experiment.

An important lesson to be drawn from all of this is the need for a clear
presentation of consistent principles of quantum reasoning in textbooks and
courses. When teaching courses on quantum information I always stress the fact
that there are no nonlocal influences in quantum theory, and point out that
this principle is useful to keep in mind when analyzing quantum
circuits. Unfortunately, physics students trained in traditional quantum
courses have difficulty replacing, or at least augmenting, the calculational
rules they learned by rote with a consistent probabilistic analysis of what is
going on. They may already have learned that the superluminal influences
reflected in violations of Bell's inequality cannot be used to transmit
information. But they also need to hear a simple explanation for why this is
so: such influences do not exist.  They are nothing but fudge factors needed
to correct a mistaken use of classical reasoning in the quantum domain.

\section*{Acknowledgments}

  I have benefitted from correspondence with G.\ Blaylock and T.\ Maudlin, as
  well as comments by anonymous referees. Support for this research has come
  from the National Science Foundation through Grant 0757251.

\end{document}